\documentclass[preprint2]{aastex}

\shorttitle{A LARGE ISOTROPIC EMISSION REVEALED BY {\it COBE}/DIRBE}

\shortauthors{Sano et al.}

\begin{document}


\title{DERIVATION OF A LARGE ISOTROPIC DIFFUSE SKY EMISSION COMPONENT AT $1.25$ AND $2.2\,\rm{\mu m}$ FROM THE {\it COBE}/DIRBE DATA}


\author{K. SANO\altaffilmark{1,2}, K. KAWARA\altaffilmark{3}, S. MATSUURA\altaffilmark{4}, H. KATAZA\altaffilmark{1,2}, \\
T. ARAI\altaffilmark{5}, AND Y. MATSUOKA\altaffilmark{6}}
\affil{
\altaffilmark{1}Department of Astronomy, Graduate School of Science, The University of Tokyo, \\
Hongo 7-3-1, Bunkyo-ku, Tokyo 113-0033, Japan\\
\altaffilmark{2}Institute of Space and Astronautical Science, Japan Aerospace Exploration Agency,\\
3-1-1 Yoshinodai, Chuo-ku, Sagamihara, Kanagawa 252-5210, Japan\\
\altaffilmark{3}Institute of Astronomy, University of Tokyo, 2-21-1, Osawa, Mitaka, Tokyo 181-0015, Japan \\
\altaffilmark{4}Department of Physics, School of Science and Engineering, Kwansei Gakuin University, 2-1 Gakuen, Sanda, Hyogo 669-1337, Japan \\
\altaffilmark{5}Frontier Research Institute for Interdisciplinary Science, Tohoku University, Sendai 980-8578, Japan \\
\altaffilmark{6}National Astronomical Observatory of Japan, 2-21-1 Osawa, Mitaka, Tokyo 181-8588, Japan \\
}

\email{sano@ir.isas.jaxa.jp}

%\and

%\author{R. J. Hanisch\altaffilmark{5}}
%\affil{Space Telescope Science Institute, Baltimore, MD 21218}


%\altaffiltext{1}{Visiting Astronomer, Cerro Tololo Inter-American Observatory.
%CTIO is operated by AURA, Inc.\ under contract to the National Science
%\altaffiltext{2}{Society of Fellows, Harvard University.}
%\altaffiltext{3}{present address: Center for Astrophysics,
 %   60 Garden Street, Cambridge, MA 02138}
%\altaffiltext{4}{Visiting Programmer, Space Telescope Science Institute}
%\altaffiltext{5}{Patron, Alonso's Bar and Grill}


\begin{abstract}
Using all-sky maps obtained with {\it COBE}/DIRBE, we reanalyzed the diffuse sky brightness at $1.25$ and $2.2\,\rm{\mu m}$, which consists of zodiacal light, diffuse Galactic light (DGL), integrated starlight (ISL), and isotropic emission including the extragalactic background light.
%To measure the extragalactic background light (EBL) at $1.25$ and $2.2\,\rm{\mu m}$, all-sky maps created by the Diffuse Infrared Background Experiment (DIRBE) aboard the Cosmic Background Explorer spacecraft were analyzed.
%At these wavelengths, the sky brightness is assumed as a linear combination of zodiacal light (ZL), diffuse Galactic light (DGL), integrated starlight (ISL), and isotropic emission including the EBL.
%Our new analysis includes an improved estimate of the DGL and the ISL by the 2MASS.
%The foregrounds were decomposed from the total sky brightness measured at high Galactic latitudes by a
%$\it{\chi^{\rm 2}}$ minimum analysis assuming linear correlations of the ZL, DGL, and ISL to the DIRBE ZL model, interstellar $100\,\rm{\mu m}$ emission, and the Two Micron All-Sky Survey Point Source Catalog, respectively.
%The intensity of the diffuse near-infrared light positively linearly correlated with the $100\,\rm{\mu m}$ emission, indicating significant DGL in the DIRBE data at $1.25$ and $2.2\,\rm{\mu m}$.
Our new analysis including an improved estimate of the DGL and the ISL with the 2MASS data showed that deviations of the isotropic emission from isotropy were less than $10\%$ in the entire sky at high Galactic latitude ($|{\it b}| > 35^\circ$).
The result of our analysis revealed a significantly large isotropic component at $1.25$ and $2.2\,\rm{\mu m}$ with intensities of $60.15 \pm 16.14$ and $27.68 \pm 6.21 \,\rm{nWm^{-2}sr^{-1}}$, respectively.
This intensity is larger than the integrated galaxy light, upper limits from $\gamma$-ray observation, and potential contribution from exotic sources (i.e., Population III stars, intrahalo light, direct collapse black holes, and dark stars).
We therefore conclude that the excess light may originate from the local universe; the Milky Way and/or the solar system.
%We also confirmed previous results that the isotropic emission, which may be attributed to the EBL, is several times brighter than the integrated light of galaxies. 

%{\bf Since the derived EBL intensity, especially at $1.25\,\rm{\mu m}$, was inconsistent with most of the $\gamma$-ray constraints on the EBL and exceeded the total contribution of the integrated light of galaxies plus currently suggested extragalactic sources (i.e., Population III stars, intrahalo light, direct collapse black holes, and dark stars), the excess light may originate from the local universe; the Milky Way and/or the solar system.}

\end{abstract}





\section{INTRODUCTION}

\subsection{Extragalactic Background Light in the Near-Infrared}

Extragalactic Background Light (EBL) in the near-infrared (IR) supposedly comprises integrated light emitted from galaxies, quasars, and possible particle decay.
Hence, the near-IR EBL is a potentially important physical indicator of star formation history and the unknown radiation processes throughout the history of the universe.

The lower limit of the near-IR EBL is the brightness of the integrated galaxy light (IGL) derived from the deep galaxy counts, such as those detected in the {\it Hubble} Deep Field (HDF) by Madau \& Pozzetti (2000) or the Subaru Deep Field (SDF) by Totani et al. (2001).
On the other hand, the upper limit of the EBL is estimated from observations of high-energy $\gamma$-ray sources, assuming the property of intrinsic spectra of the objects (e.g., Aharonian et al. 2006, Albert et al. 2008, Meyer et al. 2012).
These $\gamma$-rays interact with the EBL photons to create positron-electron pairs.
At 1--$2\,\rm{\mu m}$, the results of these two methods converge to the same intensity (i.e.,  $\sim$ 10--$20\, \rm{nWm^{-2}sr^{-1}}$).

Direct measurement of the EBL is hampered by the intense foreground emission, contributed by airglow, zodiacal light (ZL), and integrated starlight (ISL). 
In previous studies, the ZL and the ISL were subtracted from the sky brightness measured by the space telescope.  
According to Matsumoto et al. (2005), who reported on Infrared Telescope in Space ({\it IRTS}), and Wright \& Reese (2000), Cambr\'esy et al. (2001), and Levenson et al. (2007), who analyzed the Diffuse Infrared Background Experiment (DIRBE) aboard the Cosmic Background Explorer ({\it COBE}) satellite, the intensity of the EBL in the near-IR can be several times higher than that of the IGL and that estimated from the high-energy $\gamma$-ray observations.
If these findings are true, we require additional sources other than normal galaxies.
The excess light might originate from distant Population-III (Pop-III) stars which cannot be spatially resolved by recent observations.
However, Inoue et al. (2013) calculated the theoretical contribution of light from Pop-III stars to the EBL, and showed that it is smaller than the IGL contribution by 2--3 orders of magnitude.
Therefore, if the excess emission is real, we must seek other candidate sources.

\subsection{Diffuse Galactic Light}

Previous studies have revealed that diffuse Galactic light (DGL) comprises starlight scattered off by the interstellar dust grains (e.g., Elvey \& Roach 1937, Henyey \& Greenstein 1941, van de Hulst \& de Jong 1969, Mattila 1979).
The DGL contains information on the size distribution of interstellar dust grains and the interstellar radiation field (ISRF) that illuminates them.
In the optically thin limit, the intensity of the far-IR $100\,\rm{\mu m}$ emission is expected to be proportional to the DGL intensity (Brandt \& Draine 2012).
Therefore, the DGL has been quantitatively analyzed by correlating the diffuse optical light with the far-IR emission (e.g., Laureijs et al. 1987, Guhathakurta \& Tyson 1989, Paley et al. 1991, Zagury et al. 1999, Matsuoka et al. 2011, Brandt \& Draine 2012, Ienaka et al. 2013).
Although the DGL is worthy of study, it constitutes a foreground emission in the EBL measurements, and must therefore be removed before analyzing the EBL.

Thus far, the DGL and its contribution to the total sky brightness have not been quantified in the near-IR.
Leinert et al. (1998) suggested that the near-IR DGL is limited to low Galactic latitudes ($|b| < 5^\circ$), where the dust column is sufficiently dense to enhance the intensity of the scattered light. 
In contrast, Arai et al. (2015) derived the mean DGL spectrum at 0.95--$1.65 \,\rm{\mu m}$ using the low-resolution spectrometer (LRS) on the Cosmic Infrared Background ExpeRiment (CIBER) in several local regions of high Galactic latitude ($|{\it b}|\gtrsim 30^\circ$).
Although their result is consistent with the DGL spectra modeled by Brandt \& Draine (2012), whether their result is applicable to the general wide field of the sky is not clarified.
In addition, Tsumura et al. (2013b) derived the DGL spectrum at relatively low Galactic latitudes ($5^\circ \lesssim |{\it b}| \lesssim 15^\circ$) at 1.8--$5.3 \,\rm{\mu m}$ from data collected in the low-resolution prism spectroscopy mode of the Infra-Red Camara (IRC) onboard the {\it AKARI} satellite.
To elucidate the general properties of the DGL and the isotropy of the EBL, these studies 
must be supplemented by measurements of the near-IR DGL and its contribution to the sky brightness over a wide field.

\subsection{Purpose of the Present Work}

To measure the EBL and DGL at 1--$2 \,\rm{\mu m}$, we analyze data acquired by the DIRBE aboard the {\it COBE} spacecraft.
For previous studies of the diffuse IR components, the DIRBE observed the all sky from $1.25$ to $240 \,\rm{\mu m}$ in 10 bands.
These diffuse components included the ZL, thermal emission from interstellar dust, and the EBL (Hauser et al. 1998, Kelsall et al. 1998, Arendt et al. 1998, Dwek et al. 1998).
Arendt et al. (1998) subtracted the contribution of the ISL from the sky brightness using  the DIRBE Faint Source Model (FSM), which is based on the Wainscoat et al. (1992) and Cohen (1993, 1994, 1995) ``SKY'' models.
As shown in Table 4 of Arendt et al. (1998), wherein no entries appear at $1.25$ and $2.2\, \mu \rm m$, the intensity of the $100 \,\rm{\mu m}$ emission is not correlated with diffuse light at those wavelengths.
However, according to Leinert et al. (1998), the Galactic component observed by the DIRBE in these bands undoubtedly contains a scattered light contribution.
The missing DGL may have been caused by the poor precision of the DIRBE FSM which does not reproduce the actual astrometry and photometry of Galactic stars.

Since its release, the Two Micron All-Sky Survey (2MASS) Point Source Catalog (PSC) has been used for the starlight subtraction in several measurement studies of the near-IR EBL (e.g., Wright \& Reese 2000, Gorjian et al. 2000, Cambr\'esy et al. 2001, Levenson et al. 2007).
However, because their analyses were limited to small regions of low far-IR intensity at high Galactic latitude ($|{\it b}|>40^\circ$), these authors ignored the DGL  contribution.
 
The present paper reanalyzes the all-sky map created by DIRBE for the purpose of evaluating the DGL at $1.25$ and $2.2\, \mu \rm m$ and measuring the EBL.
Calculating the contribution of the ISL collected by the 2MASS PSC over a wide field of high Galactic latitudes ($|{\it b}|>35 ^\circ$), which includes both low and high $100 \,\rm{\mu m}$ emission intensity fields, we find a positive linear correlation between the $100 \,\rm{\mu m}$ emission and the diffuse near-IR light at both $1.25$ and $2.2\, \mu \rm m$.
This means that the near-IR DGL which was ignored in previous DIRBE analyses, is extracted even at high Galactic latitudes.
In fields with small DGL components, subtracting the DGL from the isotropic emission does not remove the 
excess brightness against the IGL at $1.25$ and $2.2\, \mu \rm m$, consistent with the previous studies.

The remainder of this paper is organized as follows.
Section 2 briefly describes the analyzed DIRBE data, and Section 3 shows how we estimate the contribution of each near-IR component.
In this section, the total sky brightness is decomposed into the ZL, DGL, ISL, and isotropic emission components by a ${\it \chi}^2$ minimum analysis.
Section 4 presents the fitting and evaluates the uncertainty in each component.
Section 5, compares our fitting results with those of other studies.
A summary is presented in Section 6.

Throughout this paper, the surface brightness is expressed in $\rm{nWm^{-2}sr^{-1}}$ or $\rm{MJy\,sr^{-1}}$.
The conversion formula between these units is
\begin{equation}
\it{\nu I_{\nu}} \,(\rm{nWm^{-2}sr^{-1}}) = [3000/{\it\lambda} \,(\rm{\mu m})]\, {\it I_{\nu}}\, (\rm{MJy\,sr^{-1}}).
\end{equation}

\section{DATA; DIRBE}

%% In a manner similar to \objectname authors can provide links to dataset
%% hosted at participating data centers via the \dataset{} command.  The
%% second curly bracket argument is printed in the text while the first
%% parentheses argument serves as the valid data set identifier.  Large
%% lists of data set are best provided in a table (see Table 3 for an example).
%% Valid data set identifiers should be obtained from the data center that
%% is currently hosting the data.
%%
%% Note that AASTeX interprets everything between the curly braces in the 
%% macro as regular text, so any special characters, e.g. "#" or "_," must be 
%% preceded by a backslash. Otherwise, you will get a LaTeX error when you 
%% compile your manuscript.  Special characters do not 
%% need to be escaped in the optional, square-bracket argument.

%% In this section, we use  the \subsection command to set off
%% a subsection.  \footnote is used to insert a footnote to the text.


%% This section also includes several of the displayed math environments
%% mentioned in the Author Guide.

DIRBE was primarily designed to search for the isotropic IR EBL and to measure its energy distribution.
The cryogenic operation of DIRBE was implemented from 1989 November 24 to 1990 September 21.
During these 10 months, the sky was observed in 10 bands, from $1.25\, \mu \rm m$ to $240\, \mu \rm m$.
The DIRBE instrument was designed to make accurate absolute sky-brightness measurements, with a stray light rejection of less than $1\, \rm{nWm^{-2}sr^{-1}}$ (Magner 1987) and an absolute gain calibration uncertainty of $3.1\%$ at $1.25$ and $2.2\, \mu \rm m$ (Hauser et al. 1998).
Consequently, the all-sky maps at IR wavelengths were created with $\sim 0.7^\circ$ beam size.

Since part of the present study evaluates the scaling factor of the DIRBE ZL model (Kelsall et al. 1998, hereafter called the ``Kelsall model'') against the DIRBE data themselves, as described in subsection 3.1, we use solar elongation $({\it \epsilon}) = 90^\circ$ maps from which the ZL is not subtracted.
At each pixel, the ${\it \epsilon} = 90^\circ$ maps provide both the sky coordinates and the observation date which are needed to run the Kelsall model.
In contrast, Zodi-Subtracted Mission Average (ZSMA) maps used in the previous studies (Arendt et al. 1998, Cambr\'esy et al. 2001) provide only the sky coordinates at each pixel.
Therefore, the Kelsall model cannot be used any more in the analysis of the ZSMA maps.
For this reason, we use the ${\it \epsilon} = 90^\circ$ maps in the present analysis.

In principle, DIRBE viewed every celestial line of sight through the zodiacal cloud at $90^\circ$ solar elongation once every 6 months; that is, once or twice during the 10-month cryogenic mission.
From the ${\it \epsilon} = 90^\circ$ maps, we can obtain the IR intensity of each wavelength at each pixel by interpolating the observations made at various times at ${\it \epsilon}$ close to $90^\circ$.
In the following analysis, we use the ${\it \epsilon} = 90^\circ$ maps created through 6 months of observation, starting from 1989 January 1. 
These maps cover almost all of the sky.

Panels (a) and (a') of Figure 1 illustrate the ${\it\epsilon} = 90^\circ$ map at $1.25$ and $2.2\, \mu \rm m$, respectively, on the Mollweide projection that is reprojected from the original ``{\it COBE} Quadrilateralized Spherical Cube'' (CSC) projection adopted in DIRBE products.
The CSC projection is an approximately equal-area projection that projects the celestial sphere onto an inscribed cube.
In the DIRBE convention, each cube face is divided into $256\times 256$ pixels; thus, all-sky maps have $256^2\times 6 = 393216$ pixels.
The side of each pixel is approximately $0.32^\circ$.
The following analysis is performed on the CSC projection maps.
In this paper, we use the ${\it\epsilon} = 90^\circ$ maps and the beam profile maps, available at the DIRBE website ``lambda.gsfc.nasa.gov/product/cobe/''.
%The DIRBE products are detailed in the {\it COBE} DIRBE Explanatory Supplement (1998), available at the DIRBE website. \footnote{The DIRBE products used in the present paper are available at the website ``lambda.gsfc.nasa.gov/product/cobe/''.}

\begin{figure*}
\begin{center}
 \includegraphics[scale=0.8]{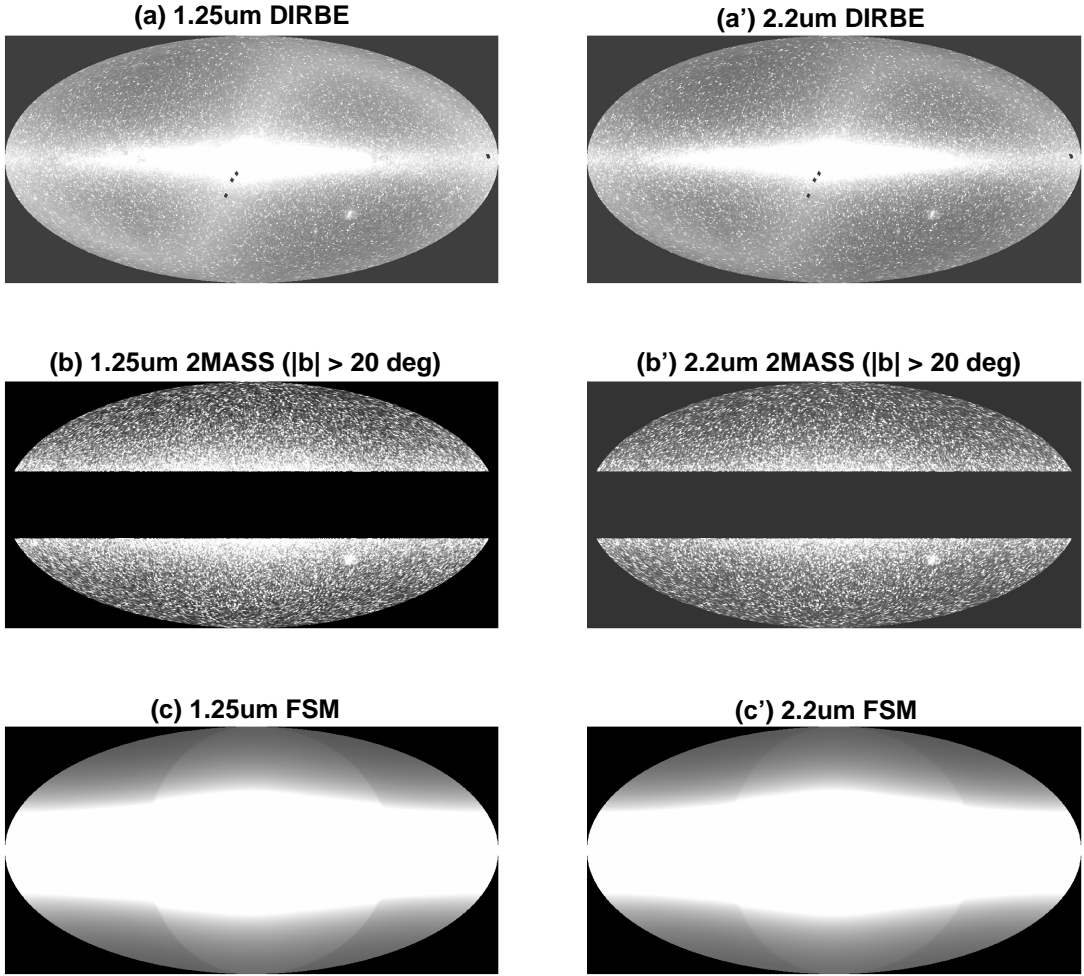} 
 \caption
 {Mollweide projections of the all-sky maps at $1.25$ (left panels) and $2.2\, \mu \rm m$ (right panels) in Galactic coordinates with the Galactic center in the middle.
 Panels (a) and (a') illustrate the full sky DIRBE ${\it \epsilon} = 90^\circ$ intensity maps.
 The ``S'' shape depicts the ecliptic plane.
Panels (b) and (b') are integrated brightness maps of the 2MASS point sources at high Galactic latitudes ($|{\it b}|>20^\circ$), created as described in subsection 3.1.3.
In the present analysis, parts of these maps ($|{\it b}|>35^\circ$) are used.
Panels (c) and (c') show the intensity maps of the DIRBE FSM used by the DIRBE team (Arendt et al. 1998), for comparison with the integrated 2MASS sources (i.e., panels (b) and (b')).
Each map is arbitrarily scaled.}

\end{center}
\end{figure*}

\section{ANALYSIS}

\subsection{Model of the Sky Brightness}

The intensity ${\it F_i(\rm{Obs})}$ of the DIRBE ${\it \epsilon} = 90^\circ$ map is evaluated by the brightness model ${\it F_i(\rm{Model})}$, where the subscript ``${\it i}$'' refers to one of two bands ($1.25$ or $2.2\, \mu \rm m$).
Within these bands, the sky brightness is assumed as a linear combination of four components: the ZL, DGL, ISL, and isotropic emission including the EBL.
The ${\it F_i({\rm Model})}$ is thus given by
\begin{equation}
{\it F_i({\rm Model}) = F_i({\rm ZL}) + F_i({\rm DGL}) + F_i({\rm ISL}) + F_i({\rm Iso})}
\end{equation}
where ${\it F_i(\rm{ZL})}$, ${\it F_i(\rm{DGL})}$, ${\it F_i(\rm{ISL})}$, and ${\it F_i(\rm{Iso})}$ denote the intensities of the ZL, DGL, ISL, and isotropic emission, respectively.
These four terms are modeled as follows.

\subsubsection{Zodiacal Light}

The ZL term $F_i(ZL)$ is defined as
\begin{equation}
{\it F_i({\rm ZL}) = a_iF_i({\rm Kel})}
\end{equation}
where ${\it a_i}$ is a free parameter and ${\it F_i({\rm Kel})}$ is the ZL brightness estimated by the Kelsall model (Kelsall et al. 1998).
The Kelsall model is a parameterized physical model fitted to the time variation of the sky brightness measured by the DIRBE.
To evaluate the scaling factor of the Kelsall model versus the DIRBE data, we adopt the free parameter ${\it a_i}$.
If the Kelsall model completely reproduces the seasonal variation of the ZL brightness observed by the DIRBE, the parameter ${\it a_i}$ will equal 1.0.

\subsubsection{Diffuse Galactic Light}

In previous studies of DGL measurements in the optical and near-IR, the intensity of the $100\,\rm{\mu m}$ emission was correlated with that of the diffuse light (e.g., Matsuoka et al. 2011, Ienaka et al. 2013, Arai et al. 2015).
In the optically thin region, the extinction of the DGL is small and the correlation is reportedly linear, consistent with theoretical expectation (see Brandt \& Draine 2012).
Theoretically, a linear correlation is expected because optically thin fields dominate at high Galactic latitudes in the $1.25$ and $2.2\, \mu \rm m$ bands.

Here, we adopt the diffuse $100\,\rm{\mu m}$ emission map created by Schlegel et al. (1998), hereafter ``SFD'', which is widely used in the correlation analyses.
To match the spatial resolutions of the SFD and DIRBE maps, we apply an $8\times 8$ pixel binning to the SFD map.
The DGL term ${\it F_i({\rm DGL})}$ is then defined as
\begin{equation}
{\it F_i({\rm DGL}) = b_iF_{{\rm 100}}}
\end{equation}
where ${\it b_i}$ is a free parameter and $F_{100}$ is the interstellar $100\,\rm{\mu m}$ intensity, defined as follows;
\begin{equation}
F_{100} = \left \{
\begin{array}{ll}
F_{{\rm SFD}}-0.8\, \rm{MJy\,sr^{-1}} & \mbox{($F_{{\rm SFD}} \geq 0.8\, \rm{MJy\,sr^{-1}}$)}\\
0.0\, \rm{MJy\,sr^{-1}} & \mbox{($F_{{\rm SFD}} < 0.8\, \rm{MJy\,sr^{-1}}$)}
\end{array}
\right.
\end{equation}
In the above expression, $F_{{\rm SFD}}$ is the $100\,\rm{\mu m}$ intensity of the SFD map.
Lagache et al. (2000) reported the EBL at $100\,\rm{\mu m}$ as $0.78\,\pm\,0.21\,\rm{MJy\,sr^{-1}}$.
Matsuoka et al. (2011) correlated the intensities of the SFD map and optical diffuse light observed by {\it Pioneer 10/11}, and revealed a clear break around $0.8\,\rm{MJy\,sr^{-1}}$.
Based on these results, we assumes $0.8\,\rm{MJy\,sr^{-1}}$ for the EBL at $100\,\rm{\mu m}$, and subtract this amount from the region of $F_{{\rm SFD}} \geq 0.8\, \rm{MJy\,sr^{-1}}$ on the SFD map to obtain the $100\,\rm{\mu m}$ brightness associated with the interstellar dust.
In the region of $F_{{\rm SFD}} < 0.8\, \rm{MJy\,sr^{-1}}$, we set $F_{100} = 0.0\, \rm{MJy\,sr^{-1}}$.

\subsubsection{Integrated Starlight}

To estimate the ISL of each region, we created integrated brightness maps of the 2MASS sources. 
The 2MASS PSC contains the photometry of approximately 470,000,000 objects covering $99.998\%$ of the sky, with accurate detections below the completeness limits ${\it J} = 15.8$ and ${\it K_s} = 14.3 \,\rm{mag}$ (Skrutskie et al. 2006, Cutri et al. 2006).\footnote{The 2MASS PSC are available at the website ``http://www.ipac.caltech.edu/2mass/releases/allsky/doc/explsup.html''.}

To convert the magnitudes of the 2MASS sources into DIRBE flux densities, we require the zero magnitude. 
As discussed by Cambr\'esy et al. (2001), the difference between the filters of the DIRBE and the 2MASS is negligible compared with the other uncertainties; hence, they need not be corrected.
In addition, Levenson et al. (2007) correlated the integrated brightness of the 2MASS PSC  sources against the DIRBE intensity in 40 high Galactic latitude regions. 
They reported common zero magnitudes at $1.25$ and $2.2\, \mu \rm m$ of $1467$ and $540 \, \rm{Jy}$, respectively.
Accordingly, these values are adopted as the zero magnitudes in the following analysis.

\begin{figure*}
\begin{center}
 \includegraphics[scale=0.8]{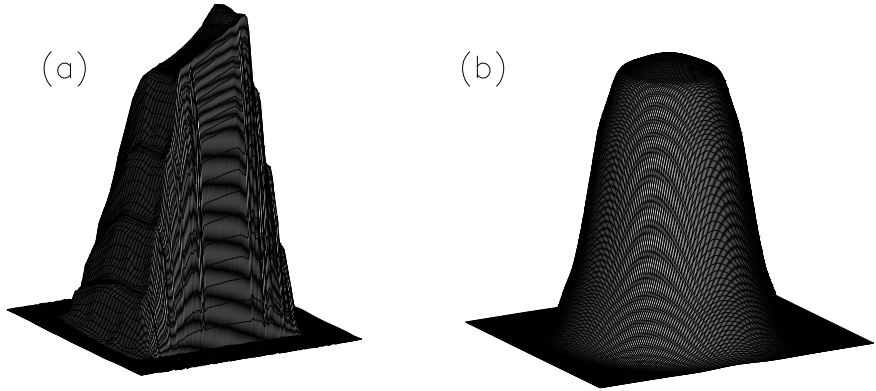} 
 \vskip -8.0cm
 \caption{Profiles of the DIRBE beams.
Panel (a) is the DIRBE beam for daily maps ($\rm{FWHM} \simeq 0.7^\circ$) at $1.25\, \mu \rm m$, available as ``Beam Profile Maps'', in the {\it COBE}/DIRBE website. 
Panel (b) is an averaged beam ($\rm{FWHM} \simeq 1^\circ$), created by averaging the effective beams of the daily maps shown in panel (a).
The integrated brightness maps in panels (b) and (b') of Figure 1 were created by applying this averaged beam to each 2MASS source.
 }
 \label{fig01}
 \end{center}
 \vskip 0.0cm
\end{figure*}

We must also apply the DIRBE beam at $1.25$ and $2.2\, \mu \rm m$ to the 2MASS sources.
An effective DIRBE beam for acquiring a daily map at $1.25\, \mu \rm m$ is illustrated in  panel (a) of Figure 2.
This beam profile measures the relative response of the DIRBE to a point source, and includes the sky scanning and data sampling effects.
In the present analysis, the averaged beam should reflect the observation period of 6 months rather than the daily beam, because the intensity of the $\epsilon = 90^\circ$ maps is the average of dozens of observations.
As illustrated in panel (b) of Figure 2, the averaged beam shapes for the $\epsilon = 90^\circ$ maps are estimated by averaging the daily beam profiles.
Similar average profiles are obtained in both bands ($1.25$ and $2.2\, \mu \rm m$), with  full width at half-maximums (FWHMs) of $\sim 1^\circ$.

Since the beam shapes should not largely depend on the location in the CSC projection sky map ({\it COBE} DIRBE Explanatory Supplement 1998), we assume that each 2MASS source isotropically transfers its flux to the nearest 13 pixels on the map according to the averaged beam shape (panel (b) of Figure 2). 
Applying this scheme to the 2MASS sources at brightnesses below the completeness limit (${\it J} < 15.8$ and ${\it K_s} < 14.3\,\rm{mag}$), we calculate the integrated brightness of each pixel at high Galactic latitudes ($|{\it b}|>20 ^\circ$).
According to the Explanatory Supplement to the 2MASS All Sky data Release and Extended Mission Products (Cutri et al. 2006), Galactic extinction below Galactic latitudes of $35 ^\circ$ renders the stellar color redder than the intrinsic color. 
Since this effect attenuate the EBL and disrupt the linear combination of the fitting model (Equation (2)), we limit the following analysis to the high Galactic latitude regions ($|{\it b}|>35 ^\circ$).
Panels (b) and (b') of Figure 1 are integrated brightness maps of the 2MASS sources at $|b|>20 ^\circ$ at $1.25$ and $2.2\, \mu \rm m$, respectively.
For comparison, the FSM map used by the DIRBE team (Arendt et al. 1998) is also shown (see panels (c) and (c') of Figure 1).
In contrast to the FSM maps, wherein the surface brightness smoothly changes across the sky, the 2MASS-derived maps show clear fluctuations reflecting the astrometry and photometry of the actual sources.

In terms of the integrated brightness of the 2MASS sources (${\it F_i({\rm 2MASS})}$; see Figure 2), the total ISL term ${\it F_i({\rm ISL})}$ is defined as
\begin{equation}
{\it F_i({\rm ISL}) = c_iF_i({\rm 2MASS})}
\end{equation}
where ${\it c_i}$ is a free parameter representing the integrated brightness of stars fainter than the limiting magnitude of the 2MASS.
This formula assumes that the integrated brightness of the fainter stars and the brighter sources below the 2MASS detection limit have the same spacial distribution.  
In previous studies using the 2MASS data for star subtraction (e.g., Cambr\'esy et al. 2001, Wright 2001), the analyzed region was sufficiently small to assume isotropic ISL of fainter stars; thus the contributions of fainter stars were subtracted by star-counts models (e.g., Jarrett, SKY model).
In contrast, the present analysis covers a wide field of the high-latitude Galactic sky, where the ISL of fainter stars and brighter sources should have the same spatial distribution.
This justifies our use of Equation (6), which is free from the uncertainties introduced by the star-counts model. 
The systematic features of this simple model are discussed in subsection 5.2.

The 2MASS PSC should contain faint galaxies that are not resolved as extended sources. 
The contributions of these faint objects should be isotropic in the sky and should be included in the EBL. 
Wright (2001) estimated that galaxies with magnitudes $K_s < 14.3\,\rm{mag}$ contribute approximately order of $0.12$ and $0.14\, \rm{nWm^{-2}sr^{-1}}$ at $1.25$ and $2.2\,\rm{\mu m}$, respectively.
To derive the EBL intensity, we apply these small corrections to the isotropic term ${\it d_i}$ after the fitting procedure.

\subsubsection{Isotropic Emission}

Since the isotropic emission is assumed to be independent of the region, the term ${\it F_i({\rm Iso})}$ is defined as
\begin{equation}
{\it F_i({\rm Iso}) = d_i}
\end{equation}
where ${\it d_i}$ is a free parameter.

\subsection{Fitting}

Now the model brightness ${\it F_i({\rm Model})}$ for ${\it F_i({\rm Obs})}$ is given by
\begin{equation}
{\it F_i({\rm Model})=F_i({\rm ZL})+F_i({\rm DGL})+F_i({\rm ISL})+F_i({\rm Iso})}
\end{equation}
\begin{equation}
= {\it a_iF_i({\rm Kel})+b_iF_{{\rm 100}}+c_iF_i({\rm 2MASS})+d_i}
\end{equation}
Prior to fitting, we should remove pixels that might perturb the analysis.
To suppress the large photometric uncertainty of bright stars, we mask the pixels around stars brighter than $J = 5$ and $K_s = 4\,\rm{mag}$ on the CSC projection map at $1.25$ and $2.2\, \mu \rm m$, respectively.
In addition, we blank out the circular regions around the Magellanic Clouds and probable Galactic extended sources listed in the Explanatory Supplement to the 2MASS All Sky Data Release and Extended Mission Products (Cutri et al. 2006).
Furthermore, we select regions with $F_{{\rm SFD}} < 10\, \rm{MJy\,sr^{-1}}$ (where the Galactic extinction is assumed negligible) and exclude outliers by applying $2$ sigma clipping to the integrated intensities of the 2MASS sources ${\it F_i(\rm{2MASS})}$.
Approximately 65\% of the total pixels in the $|{\it b}|>35 ^\circ$ region in both bands survive these masking procedures.

 To determine the parameters ${\it a_i}$, ${\it b_i}$, ${\it c_i}$, and ${\it d_i}$, we minimize the following ${\it \chi}^2$ function in each band;
\begin{equation}
{\it \chi_i^{\rm 2} = \sum_j\frac{[F_i({\rm Obs}) - F_i({\rm Model})]^{\rm 2}}{\sigma_i^{\rm 2}}}
\end{equation}
\begin{equation}
= {\it \sum_j\frac{[F_i({\rm Obs}) - a_iF_i({\rm Kel}) - b_iF_{{\rm 100}} - c_iF_i({\rm 2MASS}) - d_i]^{\rm 2}}{\sigma_i^{\rm 2}}}
\end{equation}
where ``${\it j}$'' refers to the pixels used in the fitting.
The total uncertainty ${\it \sigma_i}$ in each pixel is calculated as follows:
\begin{equation}
{\it \sigma_i^{\rm 2} = \sigma_i({\rm Obs})^{\rm 2} + b_i^{\rm 2}\sigma_{{\rm 100}}^{\rm 2} + c_i^{\rm 2}\sigma_i({\rm 2MASS})^{\rm 2}}
\end{equation}
where ${\it \sigma_i({\rm Obs})}$, $\sigma_{100}$, and $\sigma_i({\rm 2MASS})$ are the standard deviations of the intensities in the $\epsilon = 90^\circ$ map, the intensities of the $100\,\rm{\mu m}$ emission, and the integrated intensity of the 2MASS sources, respectively.
We adopt $\sigma_{100} = 0.35\, \rm{MJy\,sr^{-1}}$ derived by Ienaka et al. (2013). 
The ${\it\sigma_i ({\rm 2MASS})}$ at each pixel is calculated identically to the  brightness of the 2MASS sources (see subsection 3.1.3):
\begin{equation}
{\it \sigma_i} ({\rm 2MASS})^2 = [-0.4\,(\log{10})\,10^{-0.4 {\it m_i}}\sigma_{{\it m_i}} F_{{\it i{\rm 0}}}]^2
\end{equation}
where ${\it m_i}$, ${\it \sigma_{m_i}}$, and ${\it F_{i{\rm 0}}}$ denotes the magnitude of each 2MASS source, the uncertainty of this magnitude, and the zero magnitude derived by Levenson et al. (2007), respectively.
Sources lacking a photometric uncertainty entry in the 2MASS PSC are assigned an uncertainty of $0.5\,\rm{mag}$.

%% The equation environment wil produce a numbered display equation.

%% The \notetoeditor{TEXT} command allows the author to communicate
%% information to the copy editor.  This information will appear as a
%% footnote on the printed copy for the manuscript style file.  Nothing will
%% appear on the printed copy if the preprint or
%% preprint2 style files are used.

%% The eqnarray environment produces multi-line display math. The end of
%% each line is marked with a \\. Lines will be numbered unless the \\
%% is preceded by a \nonumber command.
%% Alignment points are marked by ampersands (&). There should be two
%% ampersands (&) per line.

%% Putting eqnarrays or equations inside the mathletters environment groups
%% the enclosed equations by letter. For instance, the eqnarray below, instead
%% of being numbered, say, (4) and (5), would be numbered (4a) and (4b).
%% LaTeX the paper and look at the output to see the results.

%% This section contains more display math examples, including unnumbered
%% equations (displaymath environment). The last paragraph includes some
%% examples of in-line math featuring a couple of the AASTeX symbol macros.

\section{RESULTS}

\begin{figure*}
\begin{center}
 \includegraphics[scale=0.7]{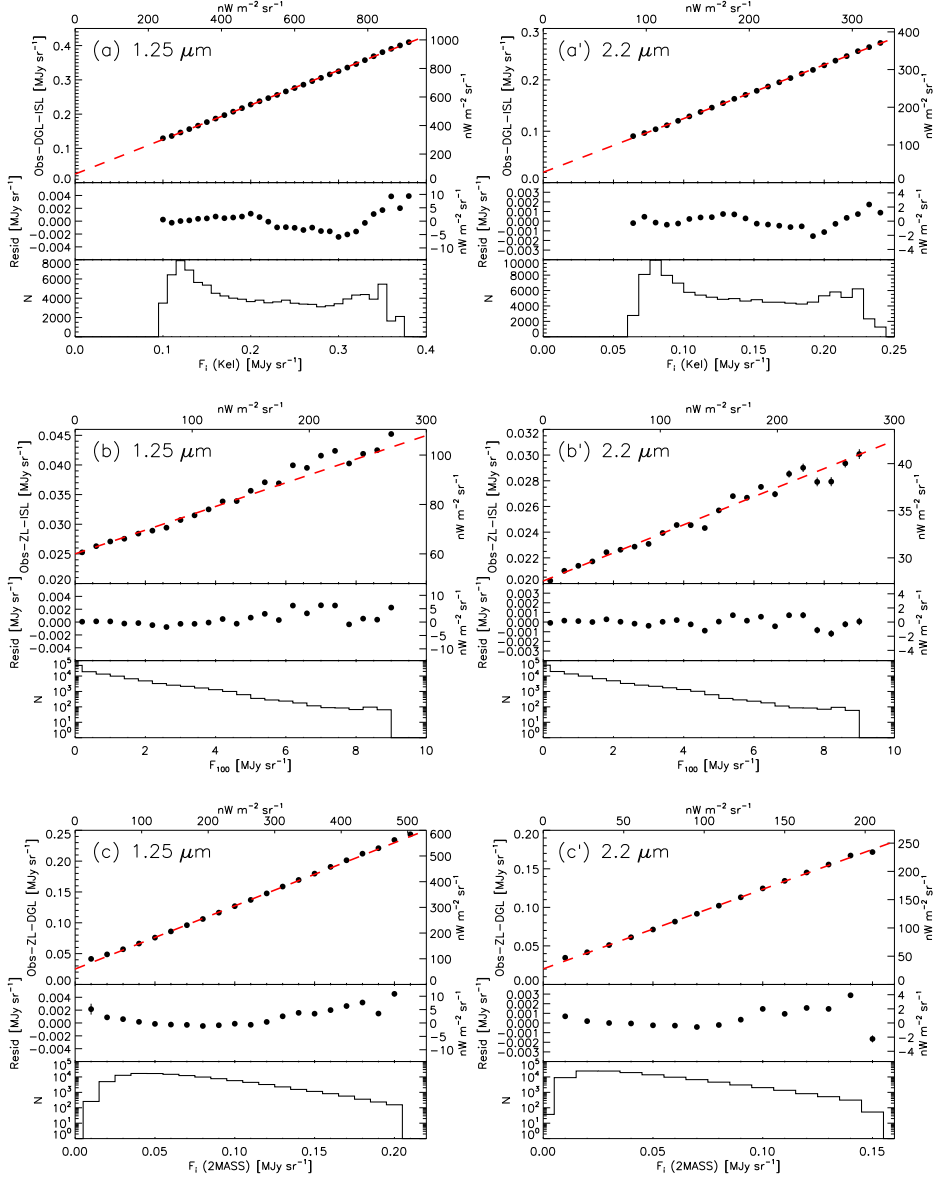} 
 \caption
 {Fitting results at $1.25$ and $2.2\,\rm{\mu m}$ ($|b|>35 ^\circ$). 
  Top panels (a) and (a') plot ${\it F_i({\rm Obs}) - F_i({\rm DGL}) - F_i({\rm ISL})}$ (i.e., ${\it a_iF_i({\rm Kel}) + d_i}$) versus ${\it F_i({\rm Kel})}$; center panels (b) and (b') plot ${\it F_i({\rm Obs}) - F_i({\rm ZL}) - F_i({\rm ISL})}$ (i.e., ${\it b_iF_{{\rm 100}} + d_i}$) versus $F_{100}$, and bottom panels (c) and (c') plot ${\it F_i ({\rm Obs}) - F_i({\rm ZL}) - F_i({\rm DGL})}$ (i.e., ${\it c_iF_i({\rm 2MASS}) + d_i}$) versus ${\it F_i({\rm 2MASS})}$.
  The red lines are the results of the best-fit parameters.
  The middle and bottom parts of each panel plot the residuals ${\it F_i({\rm Obs}) - F_i({\rm Model})}$ and the number of pixels, respectively, as functions of ${\it F_i({\rm Kel})}$ (top), $F_{100}$ (center), and ${\it F_i({\rm 2MASS})}$ (bottom).
  The filled circles and error bars represent the weighted means and the weighted standard errors of the sample within the arbitrary x-direction bin. 
 }
\label{fig04}
\end{center}
\end{figure*}

\subsection{Results of the Decomposition}

\begin{table*}[htbp]
\begin{center}
\caption{Results of the fitting in each region \label{symbols}}
 \renewcommand{\arraystretch}{1.0}
  \label{symbols}
  \scalebox{0.8}{
  \begin{tabular}{lcccccc}
  \hline
  \hline
   Band & Region & Number of & ${\it a_i}$ & ${\it \nu b_i}$ & ${\it c_i}$ &  ${\it \nu d_i}$\\
   ($\rm{\mu m}$) &  (deg)   & pixels  & (dimensionless) & $\rm{(nWm^{-2}MJy^{-1})}$ & (dimensionless) & $\rm{(nWm^{-2}sr^{-1})}$\\
   \hline  
   $1.25$ & $|{\it b}|>35$ & $116578$ & $1.0087\pm0.0001$ & $4.79\pm0.02$ & $1.0238\pm0.0003$ & $60.03\pm0.08$\\
   $2.2$  & $|{\it b}|>35$ & $119394$ & $1.0450\pm0.0002$ & $1.49\pm0.01$ & $1.0333\pm0.0004$ & $27.54\pm0.04$\\
   \hline
   \hline
   $1.25$ & $|{\it b}|>35,\, 0<{\it l}<60$ & $18557$ & $1.0017\pm0.0004$ & $8.67\pm0.06$ & $1.0206\pm0.0009$ & $64.51\pm0.25$\\
   $2.2$  & $|{\it b}|>35,\, 0<{\it l}<60$ & $18852$ & $1.0350\pm0.0006$ & $2.79\pm0.03$ & $1.0389\pm0.0011$ & $27.90\pm0.13$\\
   $1.25$ & $|{\it b}|>35,\, 60<{\it l}<120$ & $19714$ & $0.9956\pm0.0003$ & $1.91\pm0.05$ & $1.0276\pm0.0008$ & $64.25\pm0.18$\\
   $2.2$  & $|{\it b}|>35,\, 60<{\it l}<120$ & $20309$ & $1.0286\pm0.0004$ & $0.94\pm0.03$ & $1.0409\pm0.0010$ & $29.15\pm0.10$\\
   $1.25$ & $|{\it b}|>35,\, 120<{\it l}<180$ & $19582$ & $0.9993\pm0.0003$ & $5.60\pm0.04$ & $1.0236\pm0.0009$ & $68.51\pm0.20$\\
   $2.2$  & $|{\it b}|>35,\, 120<{\it l}<180$ & $20213$ & $1.0534\pm0.0005$ & $1.36\pm0.02$ & $1.0288\pm0.0011$ & $27.79\pm0.10$\\
   $1.25$ & $|{\it b}|>35,\, 180<{\it l}<240$ & $20275$ & $1.0229\pm0.0003$ & $4.07\pm0.04$ & $0.9986\pm0.0008$ & $55.97\pm0.20$\\
   $2.2$  & $|{\it b}|>35,\, 180<{\it l}<240$ & $20748$ & $1.0492\pm0.0005$ & $1.26\pm0.02$ & $1.0196\pm0.0011$ & $28.88\pm0.11$\\
   $1.25$ & $|{\it b}|>35,\, 240<{\it l}<300$ & $19516$ & $1.0224\pm0.0002$ & $0.64\pm0.06$ & $1.0223\pm0.0008$ & $53.92\pm0.19$\\
   $2.2$  & $|{\it b}|>35,\, 240<{\it l}<300$ & $20011$ & $1.0593\pm0.0004$ & $-0.07\pm0.04$ & $1.0306\pm0.0010$ & $26.12\pm0.10$\\
   $1.25$ & $|{\it b}|>35,\, 300<{\it l}<360$ & $18934$ & $1.0030\pm0.0003$ & $5.58\pm0.05$ & $1.0421\pm0.0008$ & $58.62\pm0.23$\\
   $2.2$  & $|{\it b}|>35,\, 300<{\it l}<360$ & $19261$ & $1.0372\pm0.0005$ & $2.04\pm0.03$ & $1.0524\pm0.0010$ & $25.87\pm0.12$\\
    \hline
    \end{tabular}
    }
    \end{center}
    \medskip
    Note. - The symbols in the column headings are defined in Section 3.\\
    Error in each component is the statistical uncertainty derived by the fitting.
    
\end{table*}

The parameters determined by the fitting at $1.25$ and $2.2\,\rm{\mu m}$ in the $|{\it b}|>35 ^\circ$ region are summarized in Table 1.
Owing to the large sample size (over $100,000$ points), the statistical uncertainty in each parameter is very much smaller than the determined value.
In Figure 3, the sky brightness obtained by the DIRBE observations is decomposed into the ZL, DGL, ISL, and isotropic emission according to the linear combination model (Equation (2)). 
Filled circles represent the weighted means of the points within an arbitrary x-direction bin.
In further discussion, these weighted means will be assumed as representative values.

As shown in panels (b) and (b') of Figure 3, the diffuse near-IR light is positively linearly correlated with the interstellar $100\,\rm{\mu m}$ emission at both $1.25$ and $2.2\,\rm{\mu m}$, and the correlations are significant.
This indicates that the DGL component certainly exists, even at high Galactic latitudes ($|b|>35 ^\circ$).
Moreover, the linear correlation continues through the low to high $100\,\rm{\mu m}$ intensity region ($ F_{100} \lesssim 9\, \rm{MJy\,sr^{-1}}$), indicating that the $100\,\rm{\mu m}$ emission also well-tracks the DGL at these wavelengths. 
This trend was discovered owing to the wide sky coverage of the DIRBE maps, which have wide dynamic range of the $100\,\rm{\mu m}$ intensity.
In conclusion, by combining the precise star subtraction from the 2MASS PSC data with wide-field coverage of the sky, we can find the near-IR DGL at $1.25$ and $2.2\,\rm{\mu m}$ even at low column density.

As shown in panels (a), (a'), (c), and (c') of Figure 3, the ZL and ISL are also decomposed from the sky brightness with high linearity.
Therefore, the assumed linear combination model of the sky brightness is appropriate for our purpose.

Although the residuals ${\it F_i ({\rm Obs}) - F_i({\rm Model})}$ in each panel of Figure 3 appears to be functions of ${\it F_i({\rm Kel})}$, $F_{100}$, and ${\it F_i({\rm ISL})}$, they deviate from the best-fit line by no more than $\pm10$ and $\pm5 \,\rm{nWm^{-2}sr^{-1}}$ at $1.25$ and $2.2 \,\rm{\mu m}$, respectively.
As shown in Table 2, these deviations are within $\pm2\%$ of the typical DIRBE intensity ${\it F_i ({\rm Obs})}$ in both bands.
Therefore, the large sample size of high quality DIRBE data has enabled a very precise analysis.
The possible origins of the systematic features in the residuals are discussed in section 5.2.

\begin{table*}
\begin{center}
 \renewcommand{\arraystretch}{1.0}
 \caption{Typical intensity of each component determined by the fitting at $|{\it b}|>35 ^\circ$}
  \label{symbols}
  \scalebox{1.0}{
  \begin{tabular}{lcc}
  \hline
  \hline
 
   Component ($\rm{nWm^{-2}sr^{-1}}$) & $1.25 \,\rm{\mu m} $ & $2.2 \,\rm{\mu m}$ \\
   
   \hline
   ${\it F_i({\rm ZL}) = a_iF_i({\rm Kel})}$ & $544\pm199$ & $208\pm74$ \\
   ${\it F_i({\rm DGL}) = b_iF_{{\rm 100}}}$ & $4.87\pm6.06$ &  $1.50\pm1.87$ \\ 
   ${\it F_i({\rm ISL}) = c_iF_i({\rm 2MASS})}$  & $175\pm80$ &  $66.2\pm34.7$  \\
   ${\it F_i({\rm Iso}) = d_i}$ & $60.03\pm0.08$ &  $27.54\pm0.04$ \\
   \hline
   ${\it F_i({\rm Obs})}$ & $787\pm220$ &  $304\pm87$ \\
   
    \hline
    \end{tabular}
    }
    \end{center}
    \medskip
    
    Note. - Except for ${\it F_i({\rm Iso})}$, each component is represented \\
    by its average and standard deviation of the samples in the $|{\it b}|>35 ^\circ$ region, \\
    where the regions around the 2MASS sources brighter than $J$ = 5 and $K_s$ = 4 mag \\
    at $1.25$ and $2.2\,\rm{\mu m}$ are masked, respectively.
          
 \end{table*}

\subsection{Uncertainty Estimation of the Parameters}

The statistical uncertainty of the parameters determined in the minimum ${\it \chi^{\rm 2}}$ analysis is not the only uncertainty in each component.
Other uncertainties include the parameter variation among different regions, the absolute gain of the DIRBE, the uncertainty in the faint galaxies in the 2MASS PSC, and the systematic uncertainty in the Kelsall model. 

\subsubsection{Regional Parameter Variations}

To understand the parameter variation among different regions with similar dynamic ranges of each component, we divide $|{\it b}|>35 ^\circ$ region into 6 Galactic longitude fields, i.e., $0 ^\circ < {\it l }< 60 ^\circ$, $60 ^\circ < {\it l} < 120 ^\circ$, $120 ^\circ < {\it l} < 180 ^\circ$, $180 ^\circ < {\it l} < 240 ^\circ$, $240 ^\circ < {\it l} < 300 ^\circ$, and $300 ^\circ < {\it l} < 360 ^\circ$.
In each region, we apply the fitting procedure described in subsection 3.2.
The results for each region are shown in Table 1.
Because each region contains over 10,000 points, the statistical uncertainty in this analysis remains small.
Figure 4 presents the parameters obtained in the 6 regions and in the $|{\it b}|>35 ^\circ$ region as functions of Galactic longitude.
Reasonably, the parameters determined by the fitting in the 6 regions are randomly distributed around the parameters determined in the $|{\it b}|>35 ^\circ$ field, showing some degree of scatter.
The standard deviation of the parameter values in the 6 regions is adopted as a conservative uncertainty and is listed for each parameter in the row ``Scatter'' in Table 3.

\begin{figure*}
\begin{center}
 \includegraphics[scale=0.6]{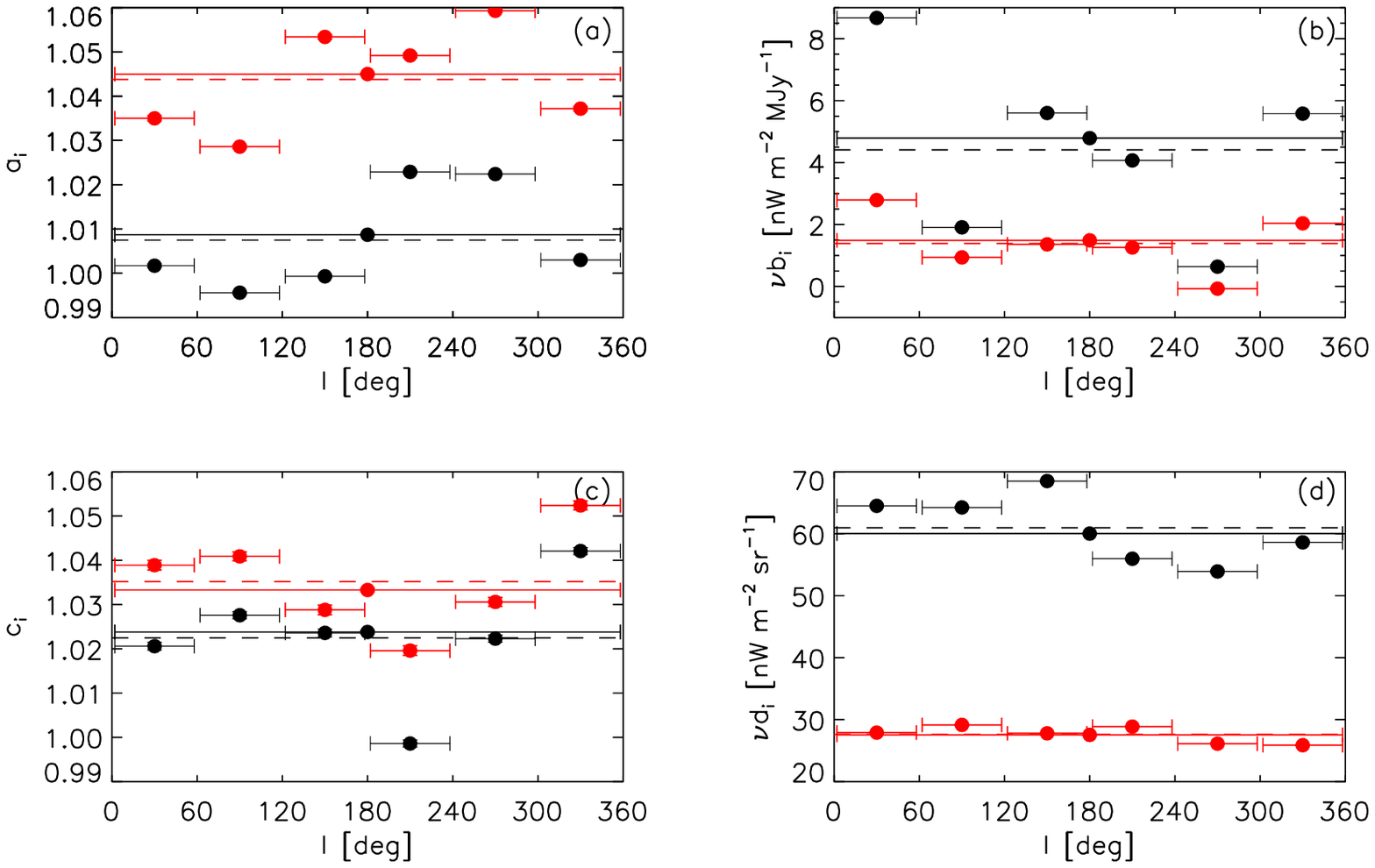} 
 \caption
 {Parameter variation among the 6 sampled regions as functions of the Galactic longitude.
 Black and red symbols are the parameter values determined at $1.25 \,\rm{\mu m}$ and $2.2 \,\rm{\mu m}$, respectively.
Panels (a), (b), (c), and (d) plot the parameters ${\it a_i}$, ${\it b_i}$, ${\it c_i}$ and ${\it d_i}$, respectively.
 Circles represent the results in each of the 6 regions and at $|{\it b}|>35 ^\circ$.
Horizontal error bars indicate the range of the region.  
Horizontal dashed lines represent the average values in the 6 regions.
}
\label{fig06}
\end{center}
\end{figure*}

\subsubsection{Uncertainty in the Absolute Gain of DIRBE}

Hauser et al. (1998) reported an uncertainty of $3.1\%$ in the absolute gain of the DIRBE at $1.25$ and $2.2 \,\rm{\mu m}$.
The Kelsall model was developed to match the photometric scale of DIRBE (Kelsall et al. 1998).
The SFD map and the ISL of the 2MASS sources are also scaled to the photometric scale of the DIRBE by the fitting process.
Then the parameters ${\it a_i}$, ${\it b_i}$, and ${\it c_i}$ are unaffected by the uncertainty in the absolute gain.
However, this uncertainty influences the parameter ${\it d_i}$.
The value of these uncertainties (assuming a percentage contribution of $3.1\%$) appear in the row ``Gain'' in Table 3.

\subsubsection{Uncertainty Associated with the Faint Galaxies in the 2MASS PSC}

As explained in subsection 3.1.3, the 2MASS PSC may contain unresolved faint galaxies in addition to the Galactic stars.
Wright (2001) estimated that galaxies with $K_s < 14.3\,\rm{mag}$ contribute around $0.12$ and $0.14\, \rm{nWm^{-2}sr^{-1}}$ to the isotropic emission at $1.25$ and $2.2\,\rm{\mu m}$, respectively.
Therefore, we add these corrections to the isotropic term ${\it d_i}$ after decomposing the integrated brightness of stars.
The contribution of unresolved galaxies is also added to the uncertainty of ${\it d_i}$ and is listed in the ``Galaxies'' row in Table 3.
Note that these contributions are relatively small.

\subsubsection{Systematic Uncertainty in the Kelsall Model}

As reported in Kelsall et al. (1998), the uncertainty in the ZL model is $15$ and $6\,\rm{nWm^{-2}sr^{-1}}$ at $1.25$ and $2.2 \,\rm{\mu m}$, respectively.
These values were estimated as the difference between two ZL models in the north Galactic pole region.
These two models were equally good in reproducing the observed seasonal variations in the ZL but not in the isotropic component.
The contributions of these uncertainties to the uncertainty in ${\it d_i}$ are listed in the row ``ZL model'' in Table 3.

\subsubsection{Total Uncertainty}

The quadrature sum of the uncertainties in each parameter is presented in the row ``Quadrature sum'' in Table 3.
In the following Discussion section, the parameters of the ZL, DGL, ISL, and isotropic emission components are assumed as the parameters determined in the $|{\it b}|>35 ^\circ$ region and their errors are assumed as the quadrature sums of the uncertainties.

\begin{table*}
\begin{center}
 \renewcommand{\arraystretch}{1.0}
 \caption{Uncertainty budget associated with each parameter}
  \label{symbols}
  \scalebox{1.0}{
  \begin{tabular}{lcccccccccccc}
  \hline
  \hline
    && \multicolumn{2}{c}{${\it a_i}$ (dimensionless)} && \multicolumn{2}{c}{${\it \nu b_i}$ ($\rm{nWm^{-2}MJy^{-1}}$)} && \multicolumn{2}{c}{${\it c_i}$ (dimensionless)} && \multicolumn{2}{c}{${\it \nu d_i}$ ($\rm{nWm^{-2}sr^{-1}}$)}\\
    \cline{3-4} \cline{6-7} \cline{9-10} \cline{12-13}
   Band ($\rm{\mu m}$) && $1.25$ & $2.2$ && $1.25$ & $2.2$ && $1.25$ & $2.2$ && $1.25 $ & $2.2$\\
 \hline
   Statistical      && 0.0001 & 0.0002  && 0.02 & 0.01 && 0.0003 & 0.0004 && 0.08 & 0.04\\
   Scatter          && 0.012 & 0.012  && 2.88 & 0.97 && 0.014 & 0.011 && 5.66 & 1.37\\ 
   Gain             && --- & --- && --- & --- && --- & --- && 1.86 & 0.85  \\
   Galaxies         && --- & --- && --- & --- && --- & --- && 0.12 & 0.14\\
   ZL model         && --- & --- && --- & --- && --- & --- && 15 & 6 \\
   Quadrature sum   && 0.012 & 0.012 && 2.88 & 0.97 && 0.014 & 0.011 && 16.14 & 6.21\\
    \hline
    Result ($|{\it b}|>35 ^\circ$)   && 1.0087 & 1.0450 && 4.79 & 1.49 && 1.0238 & 1.0333 && 60.03 & 27.54\\
    \hline
    \end{tabular}
    }
    \end{center}
    \medskip
    
    Note. - Symbols in the column headings are defined in Section 3.
    
 \end{table*}

\section{DISCUSSION}

\subsection{Interpretation of the Determined Parameters}

As shown in Table 3, the parameter ${\it a_i}$ at $1.25 \,\rm{\mu m}$ (determined as 1.0) lies within the uncertainty limits, indicating that the Kelsall model well-reproduces the time variation of the sky brightness measured by the DIRBE at $1.25 \,\rm{\mu m}$.
The parameter ${\it a_i}$ at $2.2 \,\rm{\mu m}$ exceeds 1.0 by approximately $4\%$.
On average, a $4\%$ variation in the Kelsall model corresponds to $8\,\rm{nWm^{-2}sr^{-1}}$, which is slightly larger than the claimed systematic uncertainty in the model ($6\,\rm{nWm^{-2}sr^{-1}}$).
This suggests that the Kelsall model underestimates the ZL intensity in this band.
For one thing, in the fitting procedure, the Kelsall model was sampled a sky pixel every $\sim 5^\circ$ or $\sim 10^\circ$ as a spatial grid (not used the all pixels) to avoid the excessive computational requirements.
In addition, comparing the parameter values at $1.25$ and $2.2 \,\rm{\mu m}$, some of the parameters determined in the Kelsall model, especially the phase function parameter $\it{C_{\rm 2,2}}$ and the Albedo $\it{A_{\rm 2}}$, seem anomalous $2.2 \,\rm{\mu m}$ [see Table 2 of Kelsall et al. (1998)].
Specifically, at $2.2 \,\rm{\mu m}$, $\it{C_{\rm 2,2}}$ is 3 times smaller than at $1.25 \,\rm{\mu m}$ and $\it{A_{\rm 2}}$ is unnaturally larger than that at $1.25 \,\rm{\mu m}$.
These results may change the spatial distribution of the ZL brightness in the Kelsall model, and may also explain why ${\it a_i}$ deviate from 1.0 at $2.2 \,\rm{\mu m}$.

The uncertainty in the parameter ${\it b_i}$ is dominated by scatter among the different regions and exceeds $50\%$ of the result.
This large error is attributed to the low typical brightness of the DGL (1--2 orders of magnitude fainter than the other components; Table 2).
In that situation, if the SFD intensity spatially correlates with that of the Kelsall model or the ISL to some extent, some of the DGL might be absorbed by the other components in the fitting process. 
Investigating this possibility is beyond the scope of the present study; instead, we conservatively estimate the uncertainty in the DGL as the scatter in the 6 regions.
Remarkably, the present analysis identified the DGL despite its much lower brightness than that of the other components, by virtue of the well-calibrated all-sky maps of the DIRBE.
The DGL results in the optical and near-IR, determined in the present and previous studies, are compared in subsection 5.3.

The parameter $c_i$ exceeds 1.0 by 1--4\% in both bands.
Assuming that the ISL of the sources brighter and fainter than the 2MASS detection limit have the same spatial distribution, this excess is presumably contributed by the fainter stars. 
This result also indicates that the zero magnitude derived by Levenson et al. (2007) is appropriate for converting the 2MASS magnitude to the DIRBE flux in the present study.

The determined parameter ${\it d_i}$ is discussed in section 5.4.

\subsection{Dependence of the Residuals on Galactic and Ecliptic Latitude}

Figure 5 illustrates the residuals ${\it F_i({\rm Obs}) - F_i({\rm Model})}$ derived from the fitting in the $|{\it b}|>35 ^\circ$ region as functions of Galactic latitude ${\it b}$ and ecliptic latitude ${\it \beta}$.
In general, the dependence of the residuals on Galactic latitude traces the ISL or the DGL, whose intensities are also functions of Galactic latitude.
On the other hands, the dependence on ecliptic latitude is expected to measure the accuracy of the ZL model.

\begin{figure*}
\begin{center}
 \includegraphics[scale=0.8]{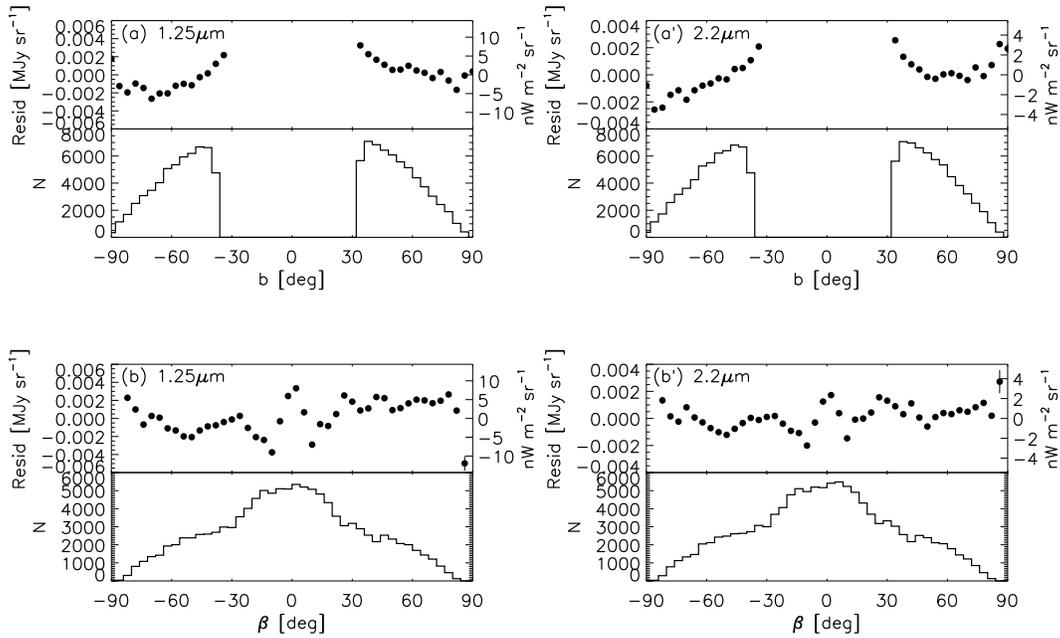} 
 \caption
 {Dependence of the residuals ${\it F_i({\rm Obs}) - F_i({\rm Model})}$ on Galactic and ecliptic latitudes, determined by the fitting in the $|{\it b}|>35 ^\circ$ region.
The upper part of each panel plots ${\it F_i({\rm Obs}) - F_i({\rm Model})}$ at $1.25$ and $2.2\,\rm{\mu m}$ as a function of Galactic latitude ${\it b}$ (top panels) and ecliptic latitude ${\it \beta}$ (bottom panels).
Filled circles and errors bars denote the weighted means and the errors of the points within arbitrary x-direction bins, respectively.
The lower part of each panel is a histogram of the number of pixels at each 
 ${\it b}$ or ${\it \beta}$.}
\label{fig04}
\end{center}
\end{figure*}

As shown in panels (a) and (a') of Figure 5, the residuals tend to increase toward low Galactic latitudes, suggesting that some component is missed at these latitudes. 
Similar trends are observed when the residuals are plotted against the integrated intensity of the 2MASS sources (panels (c) and (c') in Figure 3).
In this case, the residuals increase toward regions of higher intensity, where the data of the lower Galactic latitude fields are more dominant.

The phenomenon might stem from the contribution of stars with no entry in the 2MASS PSC, possibly because they were masked by their nearest bright sources.
According to the Explanatory Supplement to the 2MASS All-Sky Data Release and Extended Mission Products (Cutri et al. 2006), masking around bright stars can filter faint sources from the detection process.
Although the masked area in the all-sky averages to $0.25\%$ and $0.43\%$ at 1.25 and $2.2\,\rm{\mu m}$, respectively, the fraction of such regions tends to increase toward lower Galactic latitudes as the number density of bright sources increases.
In addition, the 2MASS compensated for saturation caused by bright stars by fitting the unsaturated wings of their intensity profiles.
This suggests that the 2MASS PSC could have missed faint stars.

The simply modeled ISL term ${\it F_i({\rm ISL}) = a_iF_i({\rm 2MASS})}$ might also contribute to the latitude dependence of the residuals.
If the integrated intensities of the bright and faint stars (below the detection limit of the 2MASS) have different spatial distributions, the model assumption is not strictly valid.
Although the dominant cause of the features in the residuals cannot be determined, the amplitude of the residuals is within $\pm10$ and $\pm 5\, \rm{nWm^{-2}sr^{-1}}$ at $1.25$ and $2.2\,\rm{\mu m}$, respectively.
Within these ranges, the ${\it d_i}$ terms are certainly isotropic.
The origin of the residuals' dependence on ${\it b}$ requires searching by an all-sky with superior sensitivity and spatial resolution to the 2MASS.

As illustrated in panels (b) and (b') of Figure 5, the dependence of the residuals on ecliptic latitude, especially the turbulence near the ecliptic plane, may reflect the incompleteness of the Kelsall model.
When the residuals are plotted against the Kelsall model $\it F_i({\rm Kel})$ (panels (a) and (a') of Figure 3), distortion appears in the higher intensity region (comprising fields of lower ecliptic latitudes).
Cambr\'esy et al. (2001), who similarly subtracted the ZL using the Kelsall model,  reported the same trend.
These results highlight the difficulty in applying the ZL model near the ecliptic plane, where the distribution of the zodiacal dust (including the dust bands and the circumsolar ring) becomes complex.  

The effects of these latitude dependences are naturally included in the scatter of the fitting results among the different regions (Figure 4).
Therefore, the fluctuations related to Galactic or ecliptic latitude are not added to the  uncertainty budget. 

\subsection{Spectrum of the Diffuse Galactic Light}

We now discuss the spectrum of the parameter ${\it b_i}$ in the optical and near-IR.
The four colored curves in Figure 6 are synthetic DGL spectra calculated by Brandt \& Draine (2012) based on two estimates of the ISRF continuum and two dust models; namely, the Zubko et al. (2004) and Weingartner \& Draine (2001) models, hereafter referred to as ZDA04 and WD01, respectively.
The WD01 model assumes that the half-mass grain radius $a_{0.5}$ (denoting that 50\% of the mass is contributed by grains with radii $a > a_{0.5}$) is $\sim 0.12\,\rm{\mu m}$ for both silicate and carbonaceous grains (Draine 2011).
In the ZDA04 model, grains with radii $a > 0.2\,\rm{\mu m}$ comprises a small proportion of the mass, and the half-mass radius differs between carbonaceous grains ($a_{0.5} \sim 0.06\,\rm{\mu m}$) and silicate grains ($a_{0.5} \sim 0.07\,\rm{\mu m}$). 
Consequently, the WD01 creates a redder scattered spectrum than the ZDA04 at far-optical and near-IR wavelengths.
The local ISRF continua are estimated either from Mathis et al. (1983) with de-reddening of the original ISRF of the MMP83 (see Brandt \& Draine 2012 for details) or from a synthesis model of solar-metallicity star populations (Bruzual \& Charlot 2003).
The de-reddening correction and Bruzual - Charlot model are hereafter referred to as MMP83 and BC03, respectively.
In the latter, the star formation rate is proportional to $ \exp(-t/5\, \rm{Gyr})$, where $t$ denotes the star formation timescale in units of $\rm{Gyr}$.  

\begin{figure*}
\begin{center}
 \includegraphics[scale=0.8]{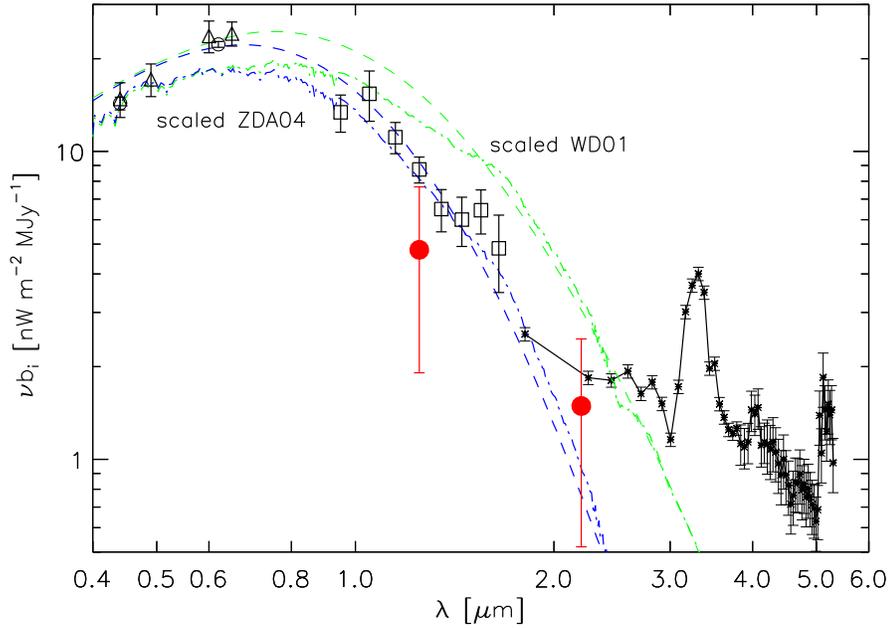} 
  \caption
 {Parameter $b_i$ as a function of wavelength in different analysis.
Plotted are the present results (filled red circles), the {\it Pioneer 10/11} data (Matsuoka et al. 2011) (open circles), results of a translucent cloud at high Galactic latitude (Ienaka et al. 2013) (open triangles), results derived from CIBER (Arai et al. 2015) (open squares), and results derived from {\it AKARI} (Tsumura et al. 2013b) (asterisks connected by the solid line). 
The colored curves are the synthetic spectra of the ratio of DGL to $100\,\rm{\mu m}$ intensity calculated by the WD01/BC03 model (green dash-dotted curve), the WD01/MMP83 model (green dashed curve), the ZDA04/BC03 model (blue dash-dotted curve), and the ZDA04/MMP83 model (blue dashed curve). 
To fit the observed results in the optical, the WD01 and ZDA04 are scaled by 1.7 and 1.9 times, respectively.
}
\label{fig06}
\end{center}
\end{figure*}

In the optical region, Figure 6 plots the results collected by {\it Pioneer 10/11} (Matsuoka et al. 2011) (open circles) and toward a high Galactic latitude translucent cloud MBM32 (Ienaka et al. 2013) (triangles).  
The {\it Pioneer 10/11} results were obtained in the same field of the sky as the present analysis ($|{\it b}| > 35^\circ$).
In the near-IR, Figure 6 plots our results (filled red circles), the mean of 6 small regions from CIBER (Arai et al. 2015) (squares), and the {\it AKARI} results (Tsumura et al. 2013b), collected at relatively low Galactic latitudes ($5^\circ < |{\it b}| < 15^\circ$) (asterisks).

Around $1.25\,\rm{\mu m}$, our result is marginally consistent with the CIBER results, considering the uncertainties in the measurements.
This suggests that results acquired in local regions represent the wider region at high Galactic latitudes.
At $2.2\,\rm{\mu m}$, our result is consistent with the {\it AKARI} results, allowing for the uncertainties. 
In contrast to our results at high Galactic latitude ($|{\it b}| > 35^\circ$), the {\it AKARI} results were taken at lower galactic latitude ($5^\circ < |{\it b}| < 15^\circ$).
Whether the relationship between the parameter ${\it b_i}$ and Galactic latitude results from the wide scatter of ${\it b_i}$ among the different regions (subsection 5.1) is difficult to determine.

The green and blue curves show the spectra of the scaled ZDA04 and WD01, respectively. 
Since the original ZDA04 and WD01 models underestimate the observed ${\it b_i}$ by a factor of 2, these models are arbitrarily scaled to the optical results (Matsuoka et al. 2011, Ienaka et al. 2013) by factors of 1.9 and 1.7, respectively.
Ienaka et al. (2013) suggested two possible explanations for this discrepancy: deficient
UV photons in the ISRF and underestimation of the assumed albedo of the dust grains in the models. 
Combining the results in the optical, CIBER, and the present study, we find that the scaled ZDA04 provides a better fitting spectrum than WD01 at $1 -2 \,\rm{\mu m}$, implying a bluing of the DGL spectrum in this wavelength range.
In contrast, the spectrum derived from {\it AKARI} significantly exceeds the ZDA04 and WD01 spectra at longer wavelengths ($\gtrsim 3 \,\rm{\mu m}$), possibly because it includes the thermal emission at low Galactic latitudes, whereas the ZDA04 and WD01 spectra include only the scattered light component.
At shorter wavelengths ($\lesssim 3 \,\rm{\mu m}$), the observed DGL can be well-fitted to the model spectra containing the scattered component alone.

{\bf \subsection{AN ISOTROPIC EMISSION COMPONENT}}

We estimate the EBL intensity from the derived parameter ${\it d_i}$.
This estimate adds the contribution of the faint galaxies appearing in the 2MASS PSC back to ${\it d_i}$. 
Assuming the faint-galaxy contribution estimated by Wright (2001) and the uncertainty in ${\it d_i}$ (Table 3), the EBL intensity at $1.25$ and $2.2 \,\rm{\mu m}$ is estimated as $60.15 \pm 16.14$ and $27.68 \pm 6.21 \,\rm{nWm^{-2}sr^{-1}}$, respectively.

\subsubsection{Isotropy test}

We now discuss the isotropy of the EBL.
Considering only the scatter of ${\it d_i}$ among the different regions (Table 3) and disregarding the other uncertainties, the deviations from isotropy are less than $10\%$ and $5\%$ of the determined ${\it d_i}$ at $1.25$ and $2.2 \,\rm{\mu m}$, respectively.
These isotropies are consistent with the typical relationships between the residuals and the Galactic and ecliptic latitudes (Figure 5).
This suggests that the EBL is isotropic within these limits even in the strong foreground emission.
The EBL isotropy can be usefully examined in the present study because of the wider observation region than in previous studies.

\subsubsection{Comparison with Other Studies}

Figure 7 compares the resultant EBL with those of previous studies.
Using the DIRBE data, Cambr\'esy et al. (2001) derived the EBL at $1.25$ and $2.2\,\rm{\mu m}$, by subtracting the Galactic stars by 2MASS and the ZL by the Kelsall model.
These authors targeted regions with low intensity of the dust emission (DIRBE $240\,\rm{\mu m}$ brightness $I_{240} < 3\,\rm{MJy\,sr^{-1}}$).
In such regions, the expected DGL brightness at $1.25$ and $2.2\,\rm{\mu m}$ is $\lesssim 7$ and $\lesssim 2 \,\rm{nWm^{-2}sr^{-1}}$, respectively, assuming the parameter ${\it b_i}$ determined in the present study and the conversion factor between the $100$ and $240\,\rm{\mu m}$ intensities ($1.297$; see Table 4 of Arendt et al. (1998)).
For this reason, Cambr\'esy et al.'s (2001) study ignored the DGL.
In addition, the $F_{100}$ histograms (panels (b) and (b') of Figure 3) show that regions of lower $100\,\rm{\mu m}$ intensity (and DGL brightness) dominate in the sky.
Therefore, the EBL results derived in this study are reasonably consistent with those obtained by Cambr\'esy et al. (2001), despite the lack of any quantitative DGL evaluation in the latter study.
However, Cambr\'esy et al. (2001) noticed fluctuations in EBL with ecliptic latitude, which they attributed to a small DGL component at $1.25\,\rm{\mu m}$ [see Figure 5 of Cambr\'esy et al. (2001)].  

Using the FSM (Faint Source Model) for the starlight subtraction, Hauser et al. (1998) derived the EBL intensity at high Galactic and ecliptic latitudes.
Although their results are plotted as 95\% confidence upper limits in Figure 7, their direct values are significantly smaller than ours; $33.0 \pm 21$ and $14.9 \pm 12 \,\rm{nWm^{-2}sr^{-1}}$ at $1.25$ and $2.2 \,\rm{\mu m}$, respectively (see Table 2 of Hauser et al. 1998).
This discrepancy can be explained by the following two things related to the ISL evaluation.
At first, in converting the magnitudes of the sources into DIRBE flux densities, the present study adopts the zero magnitude of 1467 and 540 Jy at $1.25$ and $2.2 \,\rm{\mu m}$, respectively, but Hauser et al. (1998) used higher one, i.e., 1547 and 612.3 Jy at $1.25$ and $2.2 \,\rm{\mu m}$, respectively ({\it COBE} DIRBE Explanatory Supplement 1998).
We used the zero magnitudes derived by Levenson et al. (2007), who correlated the intensity of the 2MASS-derived ISL with that of the DIRBE and corrected the zero magnitude to fit the photometric scale of the 2MASS to that of the DIRBE.
Therefore, the zero magnitudes we adopted are suitable to estimate the ISL contribution in the DIRBE data.
Next, Wright \& Reese (2000) suggested that the Wainscoat et al. (1992) star-counts model, which is the basis of the FSM, overestimates the counts by $\sim10\%$ in the $6<K<10$ range at high Galactic latitudes, compared with the 2MASS.
This is within the $10$--$15\%$ uncertainty of the FSM, estimated in Arendt et al. (1998).
Considering these differences associated with the ISL estimation, the ISL intensity in Hauser et al. (1998) can be higher than that in the present study by $\sim15$ and $\sim20\%$ at $1.25$ and $2.2 \,\rm{\mu m}$, respectively.
These percentages correspond to $\sim26$ and $\sim13 \,\rm{nWm^{-2}sr^{-1}}$ at $1.25$ and $2.2 \,\rm{\mu m}$, respectively, assuming the ISL intensity derived in the present study (Table2).
This overestimation of the ISL in Hauser et al. (1998) well explains the resultant EBL differences between Hauser et al. (1998) and the present study at both bands.

In the EBL measurement, the removal of the ZL from the sky brightness is controversial, as multiple ZL models are available.
For instance, Gorjian et al. (2000), Wright (2001) and Levenson et al. (2007) used the ZL model based on Wright (1998), whereas Cambr\'esy et al. (2001) and the present study adopted the Kelsall model.
As noted by Levenson et al. (2007), the ZL intensity at the ecliptic pole at $1.25$ and $2.2\,\rm{\mu m}$ is  $\sim 22$ and $\sim 5\,\rm{nWm^{-2}sr^{-1}}$ lower in the Kelsall model than in  Wright's (1998) model, respectively.
Consequently, the difference between the two models tends to be larger at $1.25\,\rm{\mu m}$ than at $2.2\,\rm{\mu m}$.
As shown in Figure 7, the resultant EBL obtained with the Kelsall model can be a few times lower than that obtained with the Wright model especially at $1.25\,\rm{\mu m}$.
 At $2.2\,\rm{\mu m}$, the results of both models converge within their uncertainties. 
 %Because the result remains above the IGL, an unidentified isotropic emission may also contribute to the EBL in the near-IR.
To eliminate the uncertainty introduced by the ZL model itself and its variation, we must observe outside the ZL cloud.

Note that the present decomposition analysis cannot identify where the isotropic emission including the EBL comes from.
For example, if the isotropic component associated with the ZL exists, it may contribute to the ``EBL'' called in this paper.
Actually, the Kelsall model was developed to fit to the seasonal variation of the DIRBE sky brightness, ignoring the uniform component if exists.
Hauser et al. (1998) also emphasized that the Kelsall model cannot uniquely determine the true ZL signal; in particular an arbitrary isotropic component could be added to the model.

\subsubsection{Implications of the Present Results}

Consequently, the present EBL result still remains above the observed IGL, the lower limit of the EBL, even when DGL is subtracted from the sky brightness.
%This implies that the excess is not sourced from DGL, but from other components.
As summarized in Hauser \& Dwek (2001) and Dwek \& Krennrich (2013), several studies have been modeled the intensity and the spectrum of the EBL at redshift z = 0 by different methods (e.g., Stecker et al. 2006, Mazin \& Raue 2007, Franceschini et al. 2008, Finke et al. 2010, Dom\'inguez et al. 2011).
Most of these results approach the observed IGL (Madau \& Pozzetti 2000, Totani et al. 2001, Fazio et al. 2004) and are several times lower than the present EBL results. 

To explain the excess diffuse emission reported by Matsumoto et al. (2005), the contribution of primordial (Pop-III) stars were suggested by Salvaterra \& Ferrara (2003).
However, Dwek et al. (2005b) emphasized that such a large excess is not an extragalactic origin since it would have produced a physically unrealistic intrinsic $\gamma$-ray spectrum of the blazar PKS 2155-304.
Using theoretical constraints on the formation rate of Pop-III stars, Dwek et al. (2005a) concluded that Pop-III stars can contribute only a fraction of the EBL intensity.
This is consistent with the theoretical contribution of light from Pop-III stars, i.e., $< 0.1\,\rm{nWm^{-2}sr^{-1}}$ in the near-IR (e.g., Cooray et al. 2012a, Inoue et al. 2013, Fernandez \& Zaroubi 2013).
In addition, current EBL constraints derived from the $\gamma$-ray observations, assuming the different intrinsic spectra of the sources (e.g., Dwek \& Krennrich 2005, Schroedter 2005, Aharonian et al. 2006, Mazin \& Raue 2007, Orr et al. 2011, Meyer et al. 2012), require the low EBL intensity, close to the observed IGL level.
Except that Guy et al. (2000) allowed the higher upper limit of $\sim60\,\rm{nWm^{-2}sr^{-1}}$ at $\sim1\,\rm{\mu m}$, most of the $\gamma$-ray constraints on the EBL are inconsistent with the present results at $1.25$ and $2.2\,\rm{\mu m}$.

In addition to Pop-III stars, several studies recently calculated the "exotic" sources' contribution to the EBL, such as intrahalo light (IHL), accreting direct collapse black holes (DCBH), and dark stars (DS).
The IHL could be created by tidally stripped stars from their parent galaxies by mergers and collisions (Cooray et al. 2012b). 
The IHL intensity estimated by Zemcov et al. (2014) is $\sim7$ and $\sim2\,\rm{nWm^{-2}sr^{-1}}$ 
at $1.25$ and $2.2\,\rm{\mu m}$, respectively.
Therefore, the IHL plus the observed IGL intensity approaches the EBL results derived by Wright (1998)-based ZL model (Gorjian et al. 2000, Wright 2001, Levenson et al. 2007).
Cooray et al. (2012b) and Zemcov et al. (2014) also suggested that the IHL can explain the excess in the power spectrum of the diffuse near-IR background, reported by Cooray et al. (2012a), Kashlinsky et al. (2005), Cooray et al. (2007), Thompson et al. (2007), Matsumoto et al. (2011), and  Kashlinsky et al. (2012).
To explain the excess in the power spectrum, Yue et al. (2013) suggested another candidate, i.e.,  DCBHs in the early universe.
The contribution of DCBHs to the EBL intensity has a peak at $\sim 2\,\rm{\mu m}$ and is less than $\sim1\,\rm{nWm^{-2}sr^{-1}}$ at IR wavelengths (Yue et al. 2013).
Dark stars are the hypothetical objects powered by annihilation of either accreted or captured weakly interacting massive particles before the standard nuclear fusion.
Maurer et al. (2012) separately estimated the contribution of the colder DS and the hotter ones.
As a result, the contribution of the hotter DS, marginally consistent with the current EBL observation, has a peak at $\sim2\,\rm{\mu m}$ and the intensity is $\sim10$ and $\sim20\,\rm{nWm^{-2}sr^{-1}}$ at $1.25$ and $2.2\,\rm{\mu m}$, respectively. 
The sum of these exotic sources' contribution can reach the intensity of the present result at $2.2\,\rm{\mu m}$. 
In contrast, the total of these objects contributes less than $\sim20\,\rm{nWm^{-2}sr^{-1}}$ at $1.25\,\rm{\mu m}$, indicating that the present result is approximately two times higher than the estimated contribution of the exotic sources plus the observed IGL.

In conclusion, considering the $\gamma$-ray constraints and the currently suggested extragalactic sources' contribution, it is increasingly difficult to attribute all of the derived isotropic emission (called ``EBL'' in this paper) to an extragalactic origin, especially at $1.25\,\rm{\mu m}$.
Therefore, the excess isotropic components may contain light from the local universe; the Milky Way and/or the solar system.
To identify where the excess light comes from, we need more detailed investigation on the local universe as well as extragalactic studies.

%In their analysis of {\it Pioneer 10/11}, data collected in the optical, Matsuoka et al. (2011) decomposed the observed intensity into the DGL and EBL after accurate starlight subtraction.
%On-site the {\it Pioneer 10/11} spacecraft, the ZL is expected to be sufficiently weak, so that the derived isotropic emission can be regarded as the actual EBL.
%If so, a large gap may exist between the optical and the near-IR EBL at around $1\,\rm{\mu m}$.

\begin{figure*}
\begin{center}
 \includegraphics[scale=0.8]{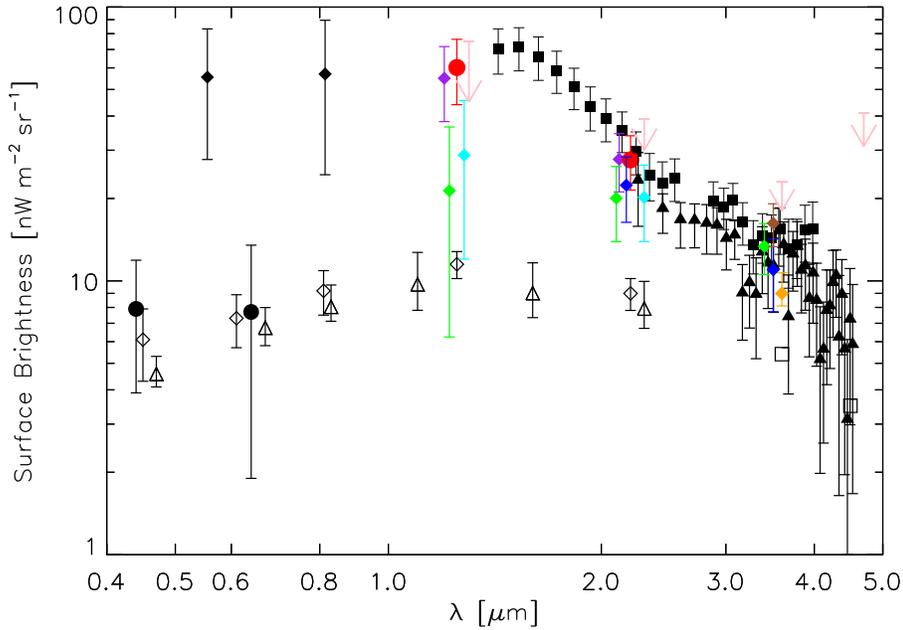} 
 \caption
 {Current measurements of the EBL in the optical and near-IR. 
  Filled red and black circles are the results of the present study and {\it Pioneer 10/11} (Matsuoka et al. 2011), respectively.
 Other EBL measurements were obtained from {\it HST}/WFPC2 (Bernstein 2007) (filled black diamonds), {\it IRTS} (Matsumoto et al. 2005) (filled squares), and {\it AKARI} (Tsumura et al. 2013a) (filled triangles).
 Colored symbols are the results of previous {\it COBE}/DIRBE data analyzed by Hauser et al. (1998) (pink down-arrows), Cambr\'esy et al. (2001) (purple diamonds), Levenson et al. (2007) (green diamonds), Gorjian et al. (2000) (blue diamonds), Wright (2001) (cyan diamonds), and Levenson \& Wright (2008) (an orange diamond). 
 A brown diamond at $3.5\,\rm{\mu m}$ is the result estimated by Dwek \& Arendt (1998), assuming the present results at $2.2\,\rm{\mu m}$.
 Open diamonds, triangles, and squares are the integrated brightness of galaxies obtained from Subaru Deep Field (Totani et al. 2001), {\it Hubble} Deep Field (Madau \& Pozzetti 2000), and {\it Spitzer}/IRAC (Fazio et al. 2004), respectively. 
 For clarity, some results are shifted a little from their exact wavelengths.
}
\label{fig07}
\end{center}
\end{figure*}

\section{SUMMARY}

We reanalyzed the {\it COBE}/DIRBE data at 1.25 and $2.2\,\rm{\mu m}$.
In particular, we measured the EBL and evaluated the DGL in the near-IR using the DIRBE data, which have wide sky coverage containing regions of both low and high interstellar $100\,\rm{\mu m}$ intensity.
To measure the contribution of the starlight at each point in the sky, the ISL intensity in each region was calculated from the 2MASS PSC, which covers almost the entire sky.

Applying a minimum $\chi^2$ analysis at high Galactic latitudes ($|{\it b}| > 35^\circ$), we decomposed the sky brightness observed by the DIRBE into its four components: ZL, DGL,  ISL, and isotropic emission.
The DGL was positively linearly correlated with the $100\,\rm{\mu m}$ brightness, confirming that the DGL exists at high Galactic latitudes in the 1.25 and $2.2\,\rm{\mu m}$ bands. 
The DGL, which is 1--2 orders of magnitude fainter than the other components, was extractable because of the high-quality wide-field data of the DIRBE.

The residuals determined by the fitting increased toward the low Galactic latitude region.
We suggested two possible causes of this phenomenon: faint stars that are filtered out from the 2MASS PSC by the nearest bright stars might remain at lower Galactic latitudes or the spatial distribution of the ISL might differ between faint and bright stars.

Previous studies have investigated the low $100\,\rm{\mu m}$ region, where the DGL contribution was found to be low in the present analysis.
After subtracting the ZL, DGL, and ISL, the intensity of isotropic emission in our study is approximately equal to that of the previous studies.
In addition, the deviations from isotropy were found to be less than $10\%$ in the entire sky at high Galactic latitudes ($|{\it b}| > 35^\circ$) at $1.25$ and $2.2\,\rm{\mu m}$.
Although the EBL intensity depends on choice of ZL models, it shows excess against the observed and expected IGL and at both investigated wavelengths.

Specifically at $1.25\,\rm{\mu m}$, the derived isotropic emission is approximately two times higher than the observed IGL plus the sum of the contribution of suggested extragalactic objects (i.e., Pop-III stars, intrahalo light, direct collapse black holes, and dark stars).
In addition, the derived isotropic emissions are larger than most of the $\gamma$-ray upper limits at both $1.25$ and $2.2\,\rm{\mu m}$.  
%This suggests that an unidentified isotropic component contributes to the EBL at $1.25$ and $2.2\,\rm{\mu m}$.
Therefore, it is possible that the excess emission originates from the local universe; Milky Way and/or the solar system.

The ISL maps created from the 2MASS PSC at $|{\it b}| > 20^\circ$ are available in the online version of this journal.

%% The displaymath environment will produce the same sort of equation as
%% the equation environment, except that the equation will not be numbered
%% by LaTeX.


%% Included in this acknowledgments section are examples of the
%% AASTeX hypertext markup commands. Use \url without the optional [HREF]
%% argument when you want to print the url directly in the text. Otherwise,
%% use either \url or \anchor, with the HREF as the first argument and the
%% text to be printed in the second.

\acknowledgments

We wish to thank T.~D. Brandt and K. Tsumura for providing their data as well as very useful comments and discussion. 
We are grateful to the anonymous referee for providing us with some useful comments.
K.S. is supported by Grant-in-Aid for Japan Society for the Promotion of Science (JSPS) Fellows. 

This publication uses the {\it COBE} data sets developed by the NASA Goddard Space Flight Center under the guidance of the {\it COBE} Science Working Group.
This publication also makes use of data products from the Two Micron All Sky Survey, which is a joint project of the University of Massachusetts and the Infrared Processing and Analysis Center/California Institute of Technology, funded by NASA and the National Science Foundation.


%% in parentheses, after the keyword, but they will not be verified.

%{\it Facilities:} \facility{Nickel}, \facility{HST (STIS)}, \facility{CXO (ASIS)}.



\appendix




\clearpage

%% the file, uncomment the \plotone, \plottwo, and
%% If you need a layout that cannot be achieved with \plotone or
%% \plottwo, you can invoke the graphicx package directly with the
%% \includegraphics command or use \plotfiddle. For more information,
%% please see the tutorial on "Using Electronic Art with AASTeX" in the
%% documentation section at the AASTeX Web site, http://aastex.aas.org/
%%
%% The examples below also include sample markup for submission of
%% supplemental electronic materials. As always, be sure to check
%% the instructions to authors for the journal you are submitting to
%% for specific submissions guidelines as they vary from

%% This example uses \plotone to include an EPS file scaled to
%% 80% of its natural size with \epsscale. Its caption
%% has been written to indicate that additional figure parts will be
%% available in the electronic journal.

\clearpage

%% Here we use \plottwo to present two versions of the same figure,
%% one in black and white for print the other in RGB color
%% for online presentation. Note that the caption indicates
%% that a color version of the figure will be available online.
%%

%% This figure uses \includegraphics to scale and rotate the still frame

%% If you are not including electonic art with your submission, you may
%% mark up your captions using the \figcaption command. See the
%% No more than seven \figcaption commands are allowed per page,
%% so if you have more than seven captions, insert a \clearpage

%% Tables should be submitted one per page, so put a \clearpage before

%% Two options are available to the author for producing tables:  the
%% deluxetable environment provided by the AASTeX package or the LaTeX
%% table environment.  Use of deluxetable is preferred.
%%

%% Three table samples follow, two marked up in the deluxetable environment,
%% one marked up as a LaTeX table.

%% In this first example, note that the \tabletypesize{}
%% command has been used to reduce the font size of the table.
%% We also use the \rotate command to rotate the table to
%% landscape orientation since it is very wide even at the
%% Note also that the \label command needs to be placed

%% This table also includes a table comment indicating that the full
%% version will be available in machine-readable format in the electronic

\clearpage

\end{document}